\documentclass[11pt]{article}
\usepackage{graphicx}
\begin{document}
\bibliographystyle{unsrt}
\def\ra{\rangle}
\def\la{\langle}
\def\aao{\hat{a}}
\def\aaot{\hat{a}^2}
\def\aco{\hat{a}^\dagger}
\def\acot{\hat{a}^{\dagger 2}}
\def\ano{\aco\aao}
\def\bao{\hat{b}}
\def\baot{\hat{b}^2}
\def\bco{\hat{b}^\dagger}
\def\bcot{\hat{b}^{\dagger 2}}
\def\bno{\bco\bao}
\def\beqn{\begin{equation}}
\def\eeqn{\end{equation}}
\def\bear{\begin{eqnarray}}
\def\eear{\end{eqnarray}}
\def\cdott{\cdot\cdot\cdot}
\def\bcen{\begin{center}}
\def\ecen{\end{center}}
\title{Entanglement of fields in coupled-cavities: effects of pumping and 
fluctuations}
\author{S. Sivakumar\\Materials Physics Division\\ 
Indira Gandhi Centre for Atomic Research\\ Kalpakkam 603 102 INDIA\\
Email: siva@igcar.gov.in\\
}
\maketitle
\begin{abstract}
A system of two coupled cavities is studied in the context of bipartite,  
continuous variable entanglement.   One of the cavities is pumped by an 
external classical source that is coupled  
quadratically, to the cavity field.  Dynamics of entanglement, 
quantified by covariance measure [Dodonov {\it et al}, Phys. Lett A {\bf 296},  
(2002) 73],  in the presence of  cavity-cavity coupling and external pumping  
is investigated.    The importance of tailoring the coupling between the 
cavities is brought out by studying the effects of  pump fluctuations on the 
entanglement.  

\end{abstract}
\newpage
\section{Introduction}\label{secI}

Quantum information processing requires entangled states
\cite{nielsen,leshouches}.  Many physical systems such as Josephson devices,  
trapped ions, NMR,  have been considered for quantum information processing. 
All these system are 
endowed with finite number of states for manipulation. 
Of late,  systems which are equipped with infinite dimensional Hilbert 
spaces have been analyzed in the context of teleportation, 
computation, etc \cite{braunsteinRMP,pati,furusawa,cvbook,devices}.   
Nonlinear optical processes, cavities in photonic bandgap materials are 
capable of producing states of light that are useful in quantum information 
processing.  A notable feature of bosonic modes is their robustness in 
retaining their entanglement in spite of thermal noise \cite{zanardi}. 
Also,  experiments with entangled photon states are of significance in 
testing the foundational aspects of quantum theory\cite{aspect}. 

  Microcavities have been extensively employed in generating 
atomic qubits, engineering of quantum states, creating polarization 
entangled photonic states to do proof-of-principle tests such as verification  
of Bell-type inequalities, entangling atoms and/or photons
\cite{vahala, raimond, bose}, 
studying the phenomenon of decoherence\cite{mabuchidoherty},  etc.  
More interestingly, coupled cavities have been thought of as a conduit  
to transfer and engineer entanglement between distant atomic qubits
\cite{cirac}, which 
would be required for distributed quantum computing. Cavities 
containing atoms (either 
positioned or on-flight) have been used in the previously mentioned 
contexts.   It is possible to use suitably prepared 
atoms to modify the cavity field so as to make the later useful
 in applications such as photonic qubits\cite{gheri,vanenk}, 
optical Josephson interferometers\cite{gerace}, etc. \\     
  
In this work, a system of two coupled single-mode cavities is considered.  The  
cavities are coupled by evanescent fields. Hence, the coupling strength 
can be  tailored by the proximity of the cavities and the permittivity  of the 
cavity walls. The field modes of the two cavities form a bipartite, bosonic, 
continuous variable (CV) system.   Because of the  coupling, the modes of the 
two cavities can get entangled. In addition to the inter-cavity coupling, 
one of the cavities is externally pumped.   The results presented here
 pertain to the case of treating the pumping field classically.      \\

The Hamiltonian of the system under consideration is 
\beqn
\hat{H}=\omega\left( \ano+\bno \right)+\lambda\left(\aco\bao+\aao\bco\right)
+\epsilon\acot+\epsilon^*\aaot.
\eeqn
The operators $\aao$ and $\aco$ are respectively the annihilation- and  
creation- 
operators of the mode ($a$-mode, hereafter) in the cavity which is externally 
pumped. Similarly, $\bao$ and 
$\bco$ are the relevant operators for the mode ($b$-mode, hereafter) in the 
other  cavity.   The strength of the coupling between the cavities  
is characterized by the parameter $\lambda$.  The magnitude of this 
parameter is related to the hopping strength of photons to hop between 
the cavities.  The term $\epsilon\acot+\epsilon^*\aaot$ 
corresponds to the contribution to the Hamiltonian from the interaction between 
the external pumping and the $a$-mode. This interaction, for instance, could 
be achieved by an active medium within the cavity.       
The magnitude of this interaction strength between the $a$-mode
and the classical driving field is given by the  parameter $\epsilon$ 
which incorporates the effects of the active medium and the external pumping.    
A similar approach is adopted to explain 
the maser action which includes many external modes\cite{louisell}. 
The external classical field affects entanglement as it delivers energy 
into the pumped cavity which supports the $a$-mode.  However, whether the 
entanglement is enhanced or decreased depends on the coupling ($\lambda$) 
between the cavities.  Firstly, the dynamics in the absence of the external 
pumping is studied.  These results are useful to understand the effects of 
pumping.\\ 
\section{Dynamics of fields: without pumping}

The Hamiltonian to study the dynamics in the absence of external 
driving corresponds to setting $\epsilon=0$ in $\hat{H}$. In that case, the  
Hamiltonian evolution is easily obtained since the commutator 
$[\ano+\bno,\aco\bao+\aao\bco]$ vanishes.  This implies that the evolution 
operator $\exp(-it\hat{H})$ can be factored as $\exp(-it(\aco\bao+\aao\bco))
\exp(-it(\ano+\bno))$.  This type of Hamiltonian is known in other contexts 
such as the symmetric beam splitter\cite{gerryknight}, the parametric  
conversion in a medium with oscillatory dielectric constant\cite{LYS} and   
has recently been used to examine intrinsic phase coherence in a 
laser\cite{pegg}.  The bilinear coupling $\lambda(\aco\bao+\aao\bco)$, between 
the modes cannot entangle them during evolution  if both the modes are in 
classical states initially\cite{kim,wang}; for instance, the $a$-mode in a 
coherent state and the $b$-mode in vacuum state.  To generate an entangled 
state from a product state, during evolution in the presence of bilinear 
coupling, at least one of the modes has to be in a nonclassical state.  
So, it is meaningful to assume the initial state of the cavities to be 
 a number state.  Since all number states, except the vacuum, are 
nonclassical,  the bilinear coupling may entangle the $a$- and $b$-
fields. It is experimentally possible to prepare the cavities such 
that $a$-field is a Fock state, say $\vert N \ra$,  and 
the $b$-field to be vacuum state.\\\\

Another consequence of the vanishing commutator 
$[\ano+\bno, \aco\bao+\aao\bco]=0$, is the existence of nontrivial, irreducible,  
invariant subspaces.  Each invariant subspace is the span of the states of the 
form $\{\vert m,N-m\ra\}$, where $m$ runs from zero to $N$. The state 
 $\vert m, N-m\ra$ of the modes in the cavities represents the situation in 
which the $a$-mode is in the number state $\vert m\ra$ and the $b$-mode is in 
$\vert N-m\ra$.  There is an invariant subspace associated with each value of 
$N$, an eigenvalue of the total number operator 
$\ano+\bno$. All the states in an invariant subspace are the eigenstates of the  
total number operator.  Further, the invariant subspaces corresponding to 
different total quantum numbers are disjoint.  Taking the initial state as 
$\vert N,0\ra$, the time-evolved state 
$\vert\psi(t)\ra=\exp(-i\hat{H}t)\vert N,0\ra$ is 
\beqn\label{ebs}
\vert\psi(t)\ra=\exp(-iN\omega t)\cos^N(\lambda t)
\sum_{n=0}^N\left[\frac{N!}
{n!(N-n)!}\right]^{1\over2}\tan^n(\lambda t)\vert N-n,n\ra.
\eeqn
This state belongs to the invariant subspace corresponding to a total 
quantum number $N$ which contains the initial state $\vert N,0\ra$.  An 
interesting feature is that the states of the form   $\vert m,N-m\ra$ are 
the only product states in the invariant subspace.  All the other states in a 
given invariant subspace are entangled\cite{siva}. The coefficients in the Fock 
state expansion are the binomial coefficients and the states defined in Eq. 
\ref{ebs} are referred as two-mode binomial states\cite{saleh}.

\begin{figure}
\includegraphics[height=10cm,width=12cm]{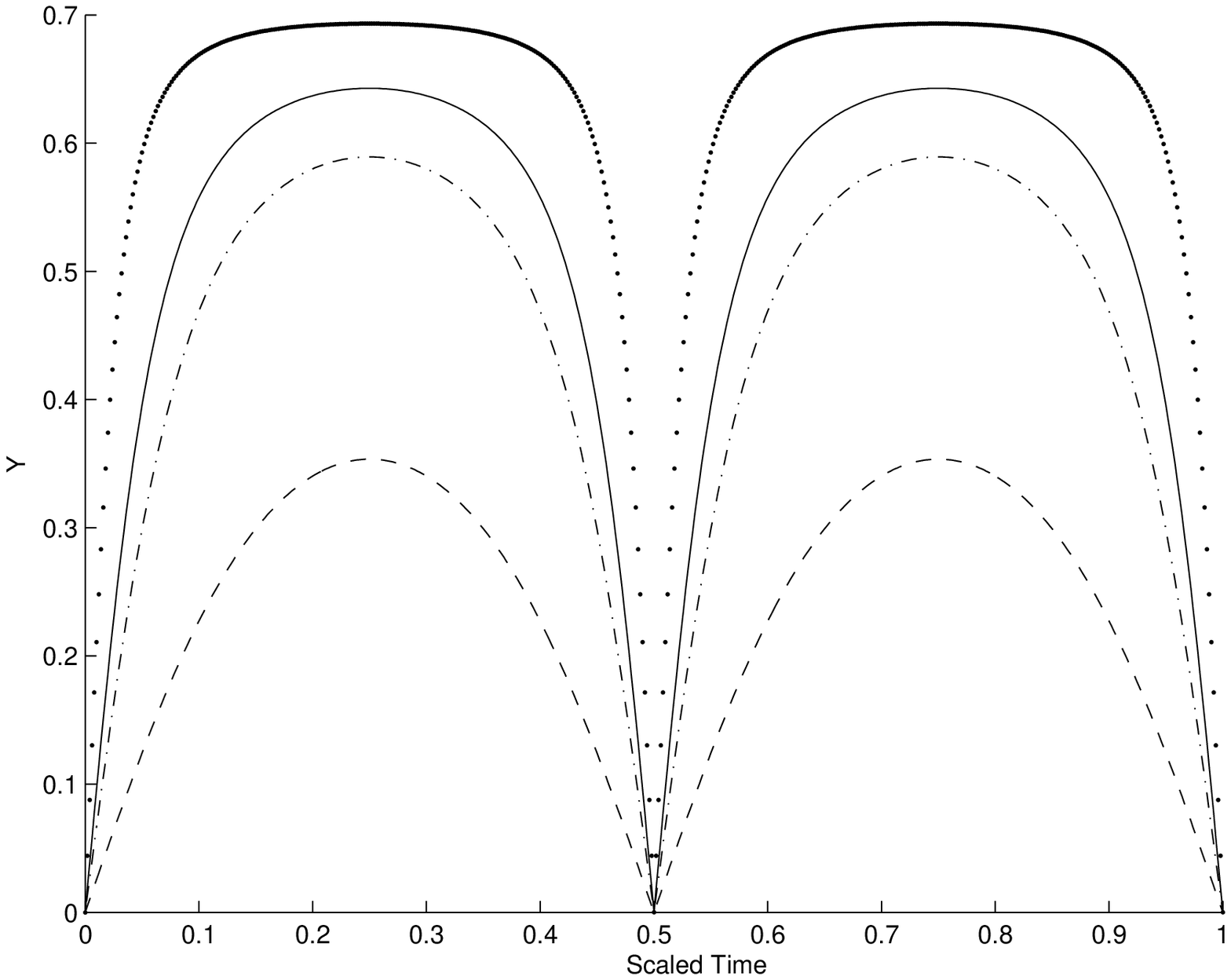}
\caption{}
\label{fig:yvst}
\end{figure}

In this work, the covariance criterion is adopted as the measure of 
entanglement\cite{dodonov}.  It is expressed as
\beqn\label{covar}
Y = \left[\frac{\vert\overline{\aao\bco}\vert^2+\vert\overline{\aao\bao}\vert^2}
{2\left(\overline{\ano}+{1\over 2}\right)\left(\overline{\bno}+{1\over 2}
\right)}\right]^{1\over 2}.
\eeqn
The bar is used to indicate covariance, for example,  $\overline{\aao\bco}=
\la\aao\bco\ra-\la\aao\ra\la\bco\ra$, where $\la\cdots\ra$ stands for  
quantum mechanical expectation value of the relevant operator.   
The quantity $Y$ is non-negative 
and less than unity.  For product states, the value of $Y$ is zero.  Nonzero 
values of $Y$ implies the state is entangled. But there exist entangled states 
for which $Y$ is zero and   hence the criterion  
is not universal.  Nevertheless, it  is easy to compute and useful in 
identifying and quantifying entanglement when it assumes nonzero values.  
If the initial state is of the form $\vert N,0\ra$, the covariance parameter of the 
evolved state given in Eq. \ref{ebs}  is 
\beqn
Y=\frac{N\vert\sin(2\lambda t)\vert}{2\left[2(N\cos^2(\lambda t)+.5)(N\sin^2
(\lambda t)+.5)\right]^{1/2}}.  
\eeqn
With nonzero coupling, 
the maximum value that $Y$ attains during evolution is $N/\sqrt{2}(N+1)$ and it 
is attained whenever $\lambda t$ is an odd integral multiple of $\pi/4$.
It is clear that the peak  value of $Y$ attained during evolution increases 
with $N$ and approaches 
$1/\sqrt{2}$ asymptotically as $N$ becomes large.  Further, the peak value 
depends on the total quantum number $N$ and not on the field-field coupling 
constant $\lambda$.
In Fig. \ref{fig:yvst} 
the variation of $Y$ with time is given for different values of the total 
quantum number, which is the eigenvalue of the operator $\ano+\bno$. The time 
of evolution is measured in the units of $\pi/\lambda$, which is the 
temporal periodicity of pumping-free evolution.  In all the subsequent 
discussions, whether the evolution is periodic or not, the time is always 
expressed in the units of $\pi/\lambda$.

\begin{figure}
\includegraphics[height=10cm,width=12cm]{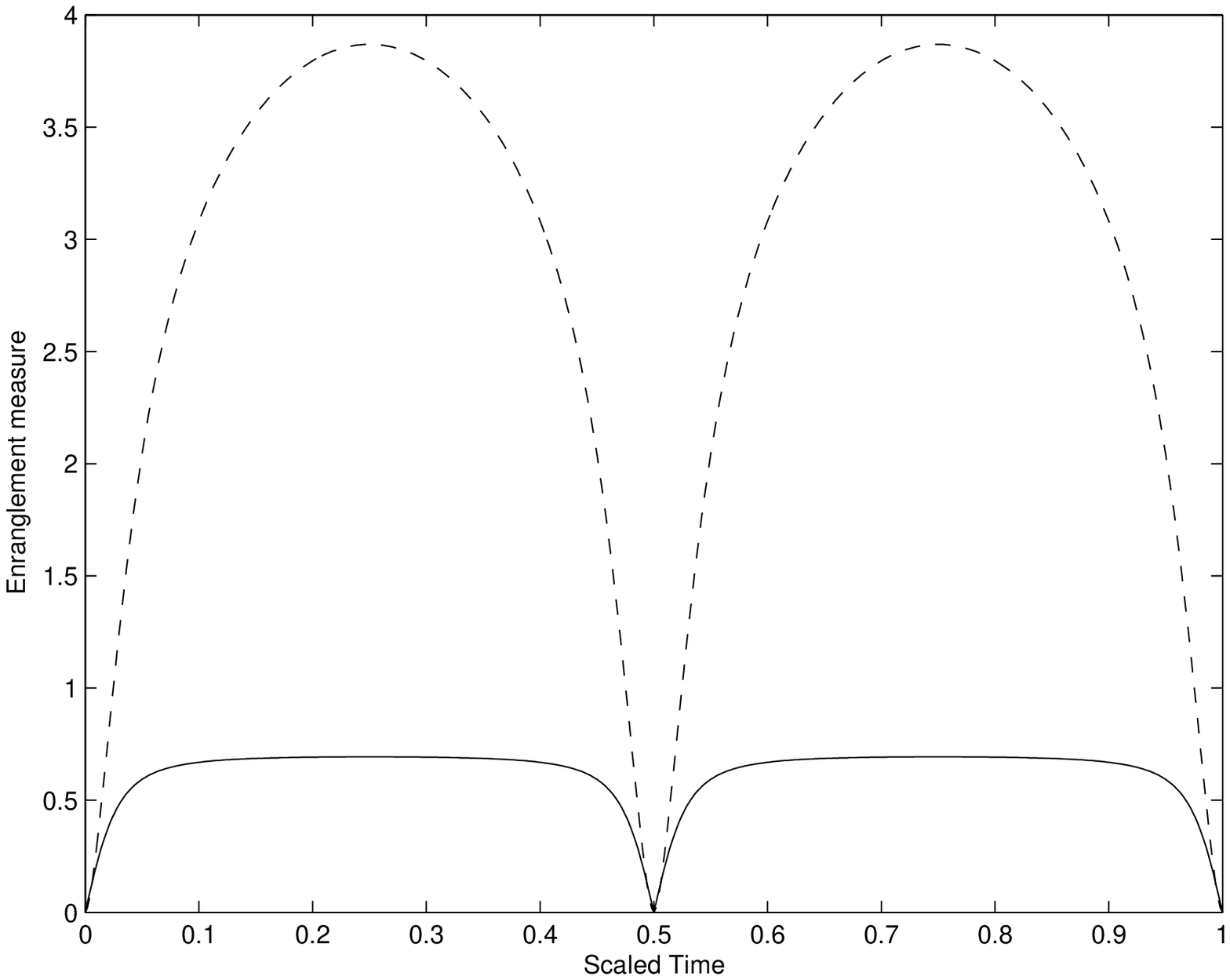}
\caption{}
\label{fig:yands}
\end{figure}

 The universal measure of entanglement for bipartite pure states\cite{bennett} 
is the von Neumann entropy $S$ defined as
\beqn
S=-\hbox{Tr}\left[\hat{\rho}_a\log_2\hat{\rho}_a\right],
\eeqn
where Tr stands for trace and $\hat{\rho}_a$ is the reduced density matrix for 
the $a$-mode field. It is instructive to  study the behaviour of von Neumann 
entropy $(S)$ {\it vis-a-vis} $Y$ in the present  case.   For the states 
defined in Eq. \ref{ebs}, the reduced density matrix $\hat{\rho}_a$ for the 
$a$-field is
\beqn\label{rdm}
\hat{\rho}_a=\cos^{2N}(\lambda t)\sum_{n=0}^N\frac{N!}{n!(N-n)!}
\tan^{2n}(\lambda t)\vert N-n\ra\la N-n\vert.
\eeqn 
Denoting the coefficient of $\vert N-n\ra\la N-n\vert$ in $\hat{\rho}_a$ by 
$c_n$, the expression for the entropy is 
$S=-2\sum_{n=0}^N\vert c_n\vert^2\log_2\vert c_n\vert$.  
In Fig. \ref{fig:yands},  the variation of $S$ and $Y$ are shown as functions 
of time taking the initial state to be $\vert 5,0\ra$.  The two measures 
exhibit very similar features.  When $S$ is nonzero, 
the covariance measure $Y$ is nonzero too, indicating that it is a good 
criterion for the kind of entangled states given in Eq. \ref{ebs}.  The peak 
in the entropy occurs when the two modes have 
nearly equal number of photons, that is, $\la\ano\ra\approx\la\bno\ra$.  Hence, 
entropy peaks correspond to minima  in 
the difference of the photon numbers of the two modes. The photon number 
in the $a$-mode is $N\cos^2(\lambda t)$ and in the $b$-mode is 
$N\sin^2(\lambda t)$.  
Therefore, $\la\ano\ra$ equals $\la\bno\ra$ whenever $\lambda t$ is an odd 
multiple of $\pi/4$.   The reason for the 
increase in the entanglement as the total number of photons increases is 
readily inferred from the expression for entropy $S$ given after Eq.\ref{rdm}.  
The entropy $S$ is maximum, subject to the constraint that 
$\hbox{Tr}\rho_a=1$,  when all $\vert c_n\vert$ are equal.  There are $N+1$ 
terms in the expression for $S$ and the maximum $S$ is $\log_2(N+1)$ 
which increases with $N$. \\


\section{Effect of external pumping}

           The evanescent coupling allows photons to hop from one cavity 
to the other.  This allows a redistribution of energy between the cavities.  
Since the $a$-mode is subjected to external pumping, the energy of the pump 
field is fed to the  $a$-mode which, in turn, is used to energize the 
$b$-field.  The pump field is treated classically and hence  the 
 driving term in the Hamiltonian is a function of the operators of the 
$a$-mode.  For weak couplings, it is reasonable to assume the 
function to be linear in $\aco$ and $\aao$. Classically, this amounts to an 
interaction proportional to the product of the field strength of the $a$- 
mode and the amplitude of the driving field.   If the 
coupling is stronger,  the function must involve higher order terms involving 
the creation and the annihilation operators.  In what follows, the effects of 
 quadratic coupling on the entanglement between the cavities 
 are discussed.  As a remark, it is noted that if the pumping is linear, that 
is, the driving term is $\epsilon\aco+\epsilon^*\aao$, there is no effect on 
the value of the covariance parameter $Y$.  There is indeed a change in the 
dynamics and, consequently the entanglement is affected.  But the dynamics of 
the covariance criterion is insensitive to such modifications, a reminiscence   
of the fact that the covariance criterion is not universal.  The quadratic 
coupling, however, leads to significant changes in the dynamics of $Y$.\\\\
  
The Heisenberg equations of motion for the operators of the two mode are 
\beqn\label{qem}
i\frac{d}{dt}
\left(
\begin{array}{c}
\aao\\
\bao\\
\aco\\
\bco
\end{array}
\right)_t
=
\left[
\begin{array}{cccc}
\omega&\lambda&2\epsilon&0\\
\lambda&\omega&0&0\\
-2\epsilon^*&0&-\omega&-\lambda\\
0&0&-\lambda&-\omega
\end{array}
\right]
\left(
\begin{array}{c}
\aao\\
\bao\\
\aco\\
\bco
\end{array}
\right)_t.
\eeqn
The equations imply that the evolution of the operators of the two modes are 
coupled.  The coefficient matrix has no explicit time-dependence and hence the  
equations are easily solved  to obtain
\beqn\label{qemsoln}
\left(
\begin{array}{c}
\aao\\
\bao\\
\aco\\
\bco
\end{array}
\right)_t
=
\exp
\left[-it\hat{M}\right]
\left(
\begin{array}{c}
\aao\\
\bao\\
\aco\\
\bco
\end{array}
\right)_0,
\eeqn
where $\hat{M}$ represents the coefficient matrix in Eq. \ref{qem}.  \\

     The operator $\exp[-it\hat{M}]$ can be expressed in terms of lower 
powers of $\hat{M}$ using Cayley-Hamilton theorem\cite{apostol}.  Let the four 
eigenvalues of $\hat{M}$ be denoted by $\alpha,\beta,\gamma$ and $\delta$. 
Defining 
$A=\omega^2+\lambda^2-2\epsilon^2$ and $B=\sqrt{\omega^2\lambda^2-\lambda^2
\epsilon^2+\epsilon^4}$, the eigenvalues are  
\bear
\alpha=-\beta=\sqrt{A-2B},\\
\gamma=-\delta=\sqrt{A+2B}.
\eear
These are the eigenvalues if $\epsilon$ is real and they are distinct if 
the determinant of $\hat{M}$ is nonzero, which is always the case for the 
typical values of the parameters assumed. 
Now,  applying 
Cayley-Hamilton theorem, 
\bear\label{expm}
4B\exp\left[-it\hat{M}\right]&=&i\left(\frac{\sin\alpha t}{\alpha}-
\frac{\sin\gamma t}{\gamma}\right)\hat{M}^3\nonumber\\
& &+\left(\cos\gamma t - \cos\alpha t\right)\hat{M}^2-
i\left(\frac{\alpha^2\sin\gamma t}{\gamma}-\frac{\gamma^2\sin\alpha t}{\alpha}
\right)\hat{M}\nonumber\\
& &-\left(\gamma^2\cos\alpha t - \alpha^2\cos\gamma t\right)I_4,
\eear
where $I_4$ is the 4$\times$4 identity matrix.\\\\

 On using the matrix exponential, refer Eq. \ref{expm}, in the solution given 
in Eq. \ref{qemsoln}, the expectation values of time-evolved operators are 
expressed  in terms of their initial expectation values.   Let the time-
dependent coefficients of $\hat{M}^3,\hat{M}^2,\hat{M}$ and $I_4$ in 
the expression for  $\exp[-it\hat{M}]$ given in Eq. \ref{expm},
be $C_3, C_2, C_1$ and $C_0$ respectively.  Further, define 
\bear
U_{\omega,\lambda}&=&C_0+\omega C_1+(A-2\epsilon^2)C_2+\omega 
(2\lambda^2-2\epsilon^2+A)C_3,\nonumber\\
V_{\omega,\lambda}&=&\lambda C_1+2\lambda\omega C_2+\lambda(2\omega^2
-2\epsilon^2+A)C_3,\nonumber\\
W_{\omega,\lambda}&=&2\epsilon C_1+ 2\epsilon(\lambda^2+A)C_3,\nonumber\\
X_{\omega,\lambda}&=&C_0+\omega C_1+(\lambda^2+\omega^2)C_2+\omega(3\lambda^2
+\omega^2)C_3,\nonumber\\
Y_{\omega,\lambda}&=&2\epsilon(\lambda C_2+\omega C_3),\nonumber\\
Z_{\omega,\lambda}&=&-2\lambda^2\epsilon C_3\nonumber.
\eear
If the initial state of the two cavities is $\vert N,0\ra$, 
the relevant quantities to compute $Y$ are 
\bear
\overline{\aao\bao}&=&(1+N)X_{\omega,\lambda}Y_{\omega,\lambda}+
NW_{\omega,\lambda}V_{\omega,\lambda}+V_{\omega,
\lambda}Z_{\omega,\lambda},\label{eab}\\
\overline{\aao\bco}&=&=(1+N)U_{\omega,\lambda}V_{-\omega,-\lambda}+
NW_{\omega,\lambda}Y_{-\omega,\lambda}+V_{\omega,\lambda}
X_{-\omega,-\lambda},\label{eabd}\\
\overline{\ano}&=&=(1+N)U_{\omega,\lambda}U_{-\omega,-\lambda}-
NW_{\omega,\lambda}^2+V_{\omega,\lambda}V_{-\omega,-\lambda}-1,\label{eaad}\\
\overline{\bno}&=&=(1+N)V_{\omega,\lambda}V_{-\omega,-\lambda}+
NY_{\omega,\lambda}Y_{-\omega,\lambda}+X_{\omega,\lambda}X_{-\omega,-\lambda}-1.\label{ebbd}
\eear

\begin{figure}
\includegraphics[height=10cm,width=12cm]{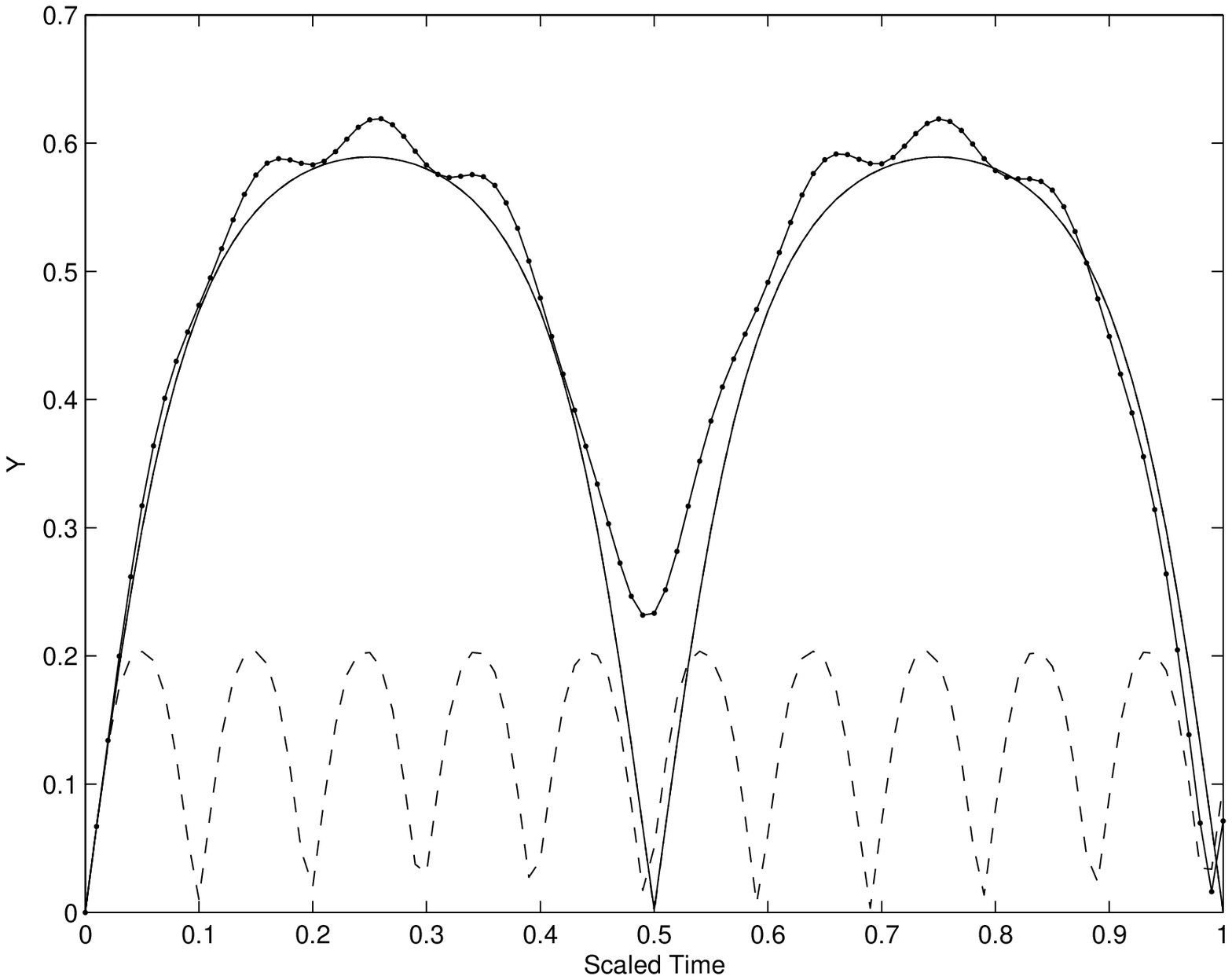}
\caption{}
\label{fig:y2pump}
\end{figure}

\begin{figure}
\includegraphics[height=10cm,width=12cm]{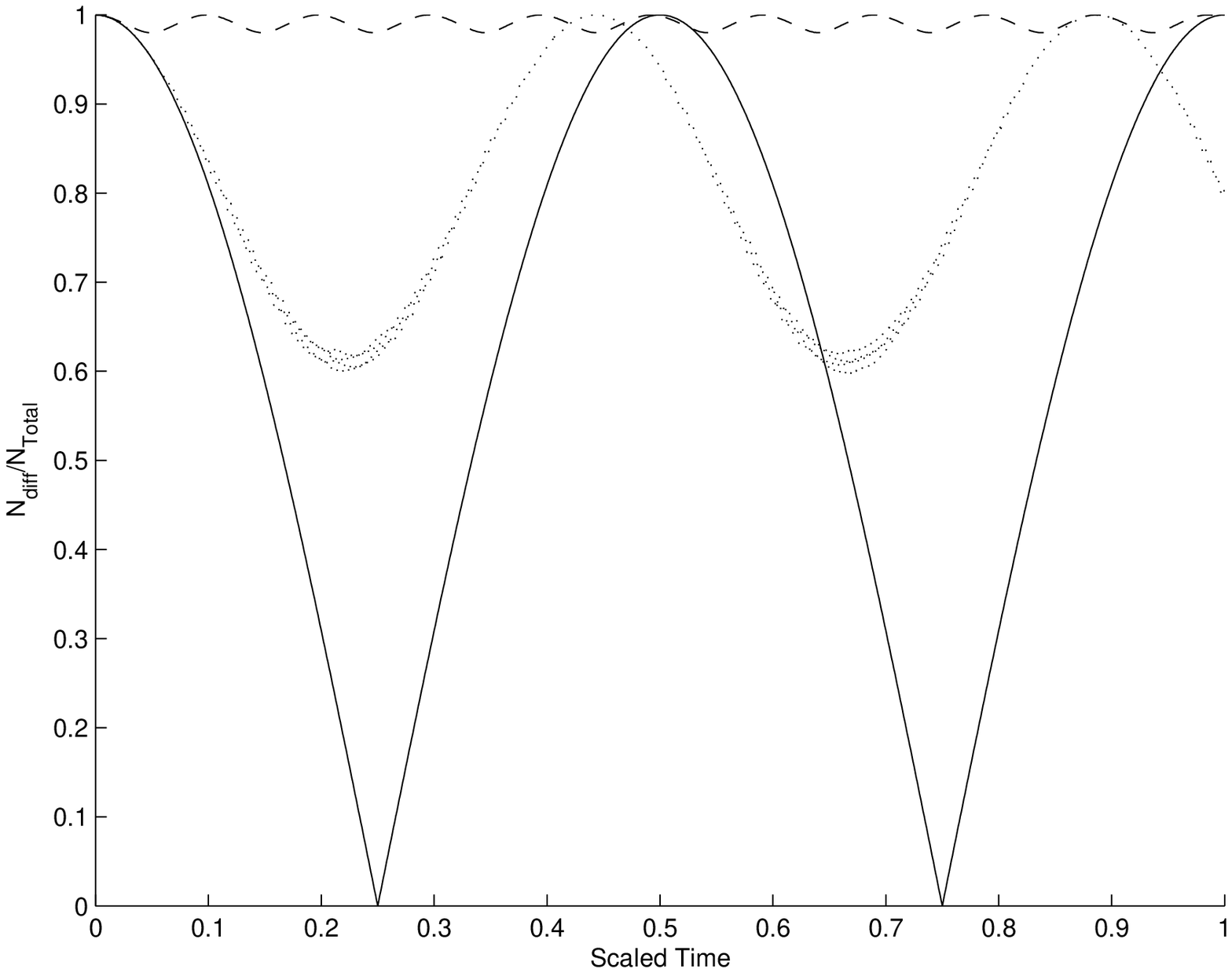}
\caption{}
\label{fig:frenergy}
\end{figure}

Using the expressions in Eqs. \ref{eab}-\ref{ebbd} for the various bilinear 
combinations of the creation- and annihilation- operators of the two cavity modes, 
the covariance criterion of entanglement is computed.  
In Fig. \ref{fig:y2pump}  the evolution of $Y$ with time is shown for various 
combinations of $\lambda$ and $\epsilon$. When both $\lambda$ and $\epsilon$  
are chosen to be 0.1, the entanglement measure $Y$  builds up to 0.6.  The 
occurrence of  peak values in $Y$ is corroborated with the occurrence  of 
minimum in the difference between the photon numbers of the two modes.  The 
variation of the ratio of the mean photon number difference 
$\vert\la\ano-\bno\ra\vert$ to the total number of photons 
$\vert\la\ano+\bno\ra\vert$ is shown in Fig. \ref{fig:frenergy}.  If the 
ratio is unity, the photon number of one of the modes is zero.  If the ratio 
is zero, the two modes have equal number of photons.  The evolution of the 
ratio is shown for three different combinations of $\lambda$ abd $\epsilon$.  
If the cavity-cavity coupling $\lambda$ is sufficiently strong that the 
energy increase due to the external pumping gets distributed to both the cavities, 
the difference in the average photon numbers in the two cavities can become 
zero, as in the case of $\lambda$ and $\epsilon$ both being 0.1.  If $\lambda$ 
is small, the energy transfer between the cavities  
is ineffective and the photon number difference is large, for instance, when
$\lambda=0.001$ and $\epsilon=0.1$.  To illustrate the effect of larger $\lambda$, 
the evolution of the ratio if $\lambda=0.005$ and $\epsilon=0.1$ is 
shown in Fig. \ref{fig:frenergy}.  In this case, the difference becomes smaller than 
the $\epsilon =0.001$ case; however, the coupling is still not strong enough 
for efficient energy exchange between the cavities to make the  difference in the 
photon number to become zero.\\\\

The occurrences of  vanishing photon-number difference and the peak in $Y$   
happen at  same instants, seen by comparing  Figs. \ref{fig:y2pump} and 
\ref{fig:frenergy}.  When the pump-field coupling is large, the $a$-mode 
gets energized at a faster rate.  If the coupling $\lambda$ is such that the 
exchange of energy between the cavities is effective, the $a$-field and the $b$-
field  attain nearly equal energies so that the difference of their photon 
numbers becomes small at various instants.  This, as in the case of pump-free 
evolution, is correlated with the occurrence of entanglement peaks. Similarly, 
if both $\lambda$ and $\epsilon$ are equal but smaller, chosen to be 0.001 for 
the purpose of discussion, the slow rate of energy build 
up in the $a$-mode is matched by the flow of energy between the modes.  
Hence, in this case too the energies of the two modes can be nearly equal 
during evolution.  At such instants, the entanglement becomes higher. \\

If $(\lambda<<\epsilon)$, for instance, $\lambda=0.001$ and $\epsilon = 0.1$, 
the external pumping primarily 
enhances the energy of the $a$-field as there is no effective flow of 
energy to the $b$-field because of low value of  the coupling 
between the cavities.  This increases the energy difference between the modes.  
The evolution of $Y$ in this case is shown in Fig. \ref{fig:y2pump}.    
Compared to the pump-free evolution of the initial state $\vert 5,0\ra$ shown 
in Fig. \ref{fig:yvst}, the maximum entanglement attained is smaller as the 
energy difference between the modes is larger.  The $b$-mode is initially 
in vacuum state. Due to the weak coupling between the cavities, the 
$b$-mode remains in a state in which only the vacuum state and a few number 
states are present.  Therefore, the entanglement becomes smaller.  The 
entanglement attains a peak whenever the energy difference is minimum.  Thus, 
generation of large entanglement requires that energy flows into the $b$-field, 
which, in turn, requires a strong interaction between the cavities.\\ 

In the opposite limit $\lambda >> \epsilon$, for instance $ 
\lambda=0.1$ and $\epsilon=0.001$, the total energy of the system does not 
increase much as pumping is weak.  The dynamics is very similar to that in the 
case when there is no pumping.  The effective coupling between the modes leads 
to nearly equal energies in the modes during evolution.  Whenever the mean 
photon numbers of the modes match, the entanglement attains a peak. The 
evolution profile of $Y$ is not distinctly seen in Fig. \ref{fig:y2pump}
as it merges with the profile corresponding to $\lambda=0.001$ and 
$\epsilon = 0.001$.\\

\section{Effect of pump fluctuations}
\begin{figure}
\includegraphics[height=10cm,width=12cm]{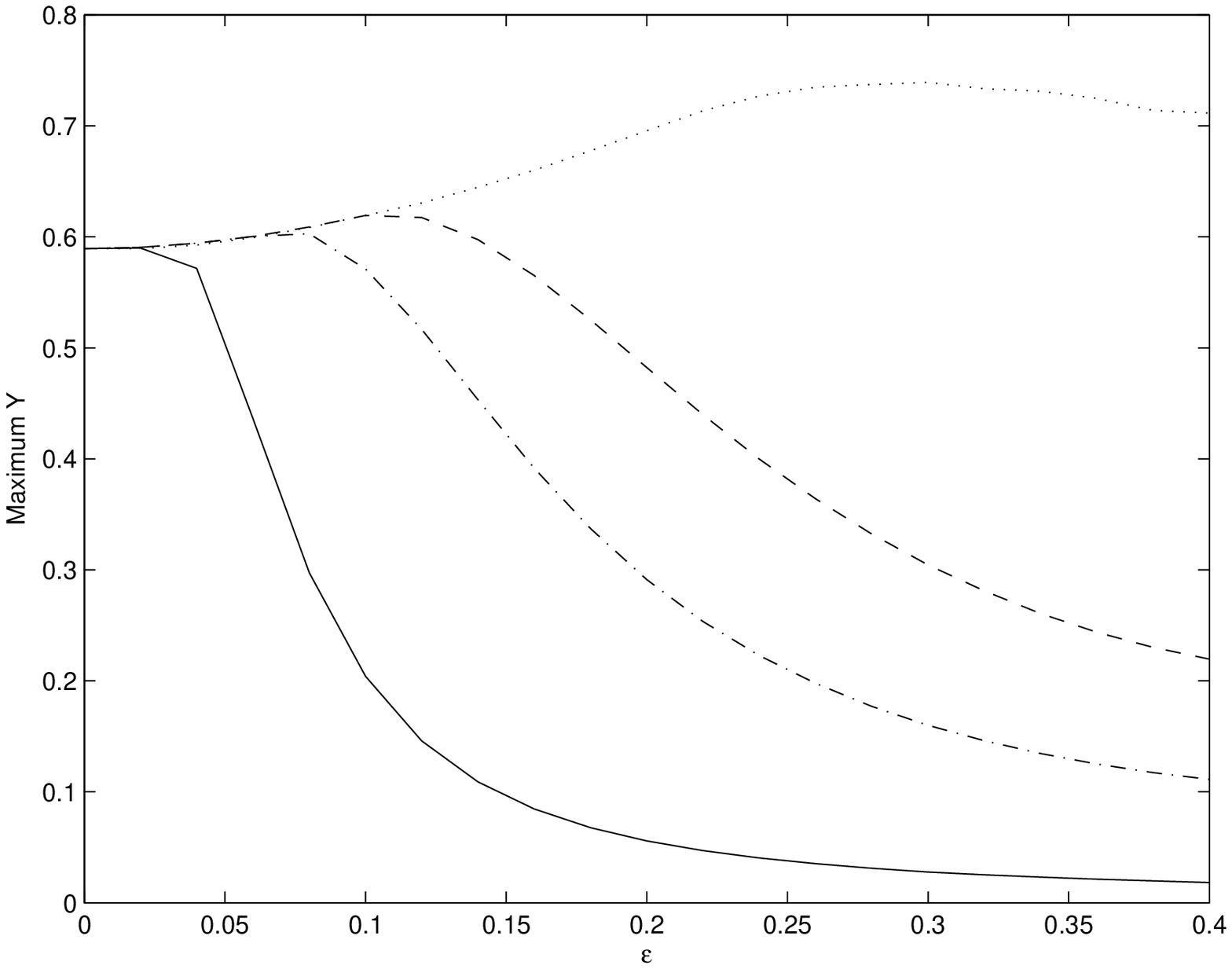}
\caption{}
\label{fig:maxy}
\end{figure}

The maximum value that $Y$ attains during evolution is dependent on the 
initial photon number $N$, the coupling $\lambda$ and the 
pump-field interaction strength $\epsilon$.  In Fig. \ref{fig:maxy}  the 
maximum value of $Y$ attained during evolution is shown  as a function of 
$\epsilon$.  Results corresponding to two different values of $\lambda$ are 
given taking the initial state to be $\vert 5,0\ra$.  It is seen that the 
peak value increases with $\epsilon$ if $\lambda$ 
is 0.1 whereas it decreases drastically if $\lambda=0.001$.  If $\lambda=0.1$,  
the peak value of $Y$ is close to 0.6, for small values of $\epsilon$.  
This is nearly the value that is  attained in the absence of pumping since 
the pumping is weak.  As the value of $\epsilon$ increases beyond 0.05, the 
peak  value increases.  For sufficiently large $\epsilon$, say, $\approx 0.4$,  
the peak value corresponds to the maximum attainable in pumping-free case and  
initial $N$ as large as 50.  This is understandable as large pumping leads 
to large energy input. Since $\lambda=0.1$ is sufficient  enough for energy 
transfer between the modes, both  the modes will have more number of Fock 
states in their superposition.  Hence, a large entanglement is possible.  If 
$\lambda=0.001$, a representative value for small $\lambda$, the coupling is 
not good enough for the energy to flow from the $a$-mode.  Thus, the $b$-mode 
will have fewer number of Fock states in the superposition. Also, as $\epsilon$ 
increases,  the difference in the photon number of  the modes will become larger.  
So, there is reduction in the maximum entanglement between the modes as compared 
with the values in the absence of pumping.   The dependence  of peak $Y$ on 
$\epsilon$ is shown in Fig. \ref{fig:maxy} for two other values of $\lambda$.  
It is seen that as $\lambda$ becomes larger, the peak $Y$ during  evolution 
is nearly constant  over a larger range of $\epsilon$.



\begin{figure}
\includegraphics[height=10cm,width=12cm]{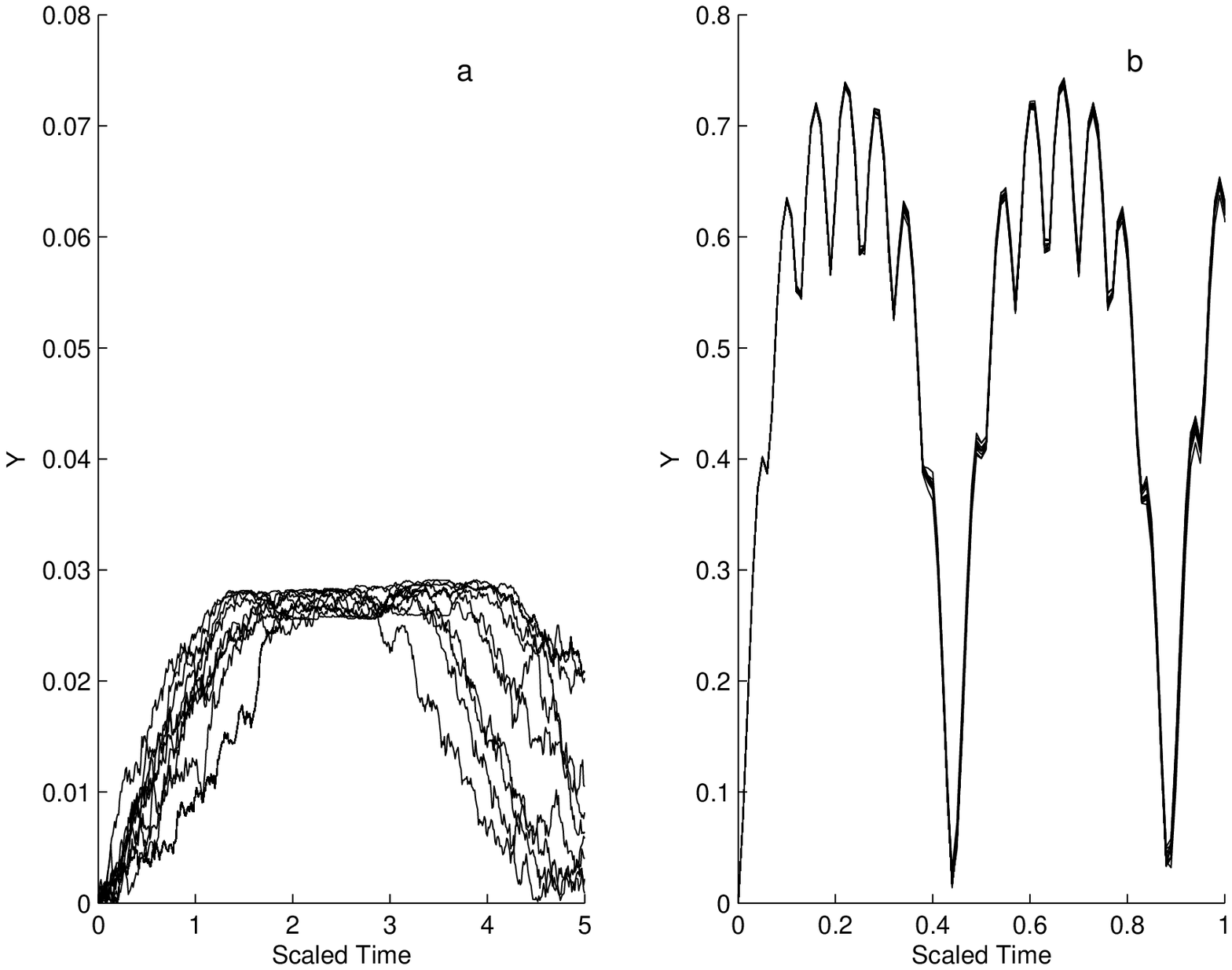}
\caption{}
\label{fig:fluc}
\end{figure}

   From Fig. \ref{fig:maxy}, it is seen that if $\lambda$ is 0.1,  
the maximum attainable $Y$ varies from 0.6 to 0.7 as $\epsilon$ increases 
from 0 to 0.5.  On the other hand, if $\lambda=0.001$ 
the maximum attainable $Y$ changes from 0.6 to 0.03 as $\epsilon$ varies 
from 0 to 0.5.  The change in maximum $Y$ is larger for the smaller 
$\lambda$ implying a larger sensitivity to changes in $\epsilon$.  This 
has important consequences.  In any pumping scheme there are fluctuations 
in the driving amplitude.  The effect of these fluctuations on the 
entanglement is decided by the sensitivity of the system.  From the 
discussion it is expected that the dynamics should not differ much if 
$\lambda$ is large whereas perceptible changes are possible if 
$\lambda$ is small.  To account for the driving field 
fluctuations, the coefficient $\epsilon$ is assumed to fluctuate about its mean 
value.   The fluctuation is assumed to be Gaussian with mean equal to the value 
of assumed value of $\epsilon$ and variance to be one-tenth of the mean.  
When the mean value of $\epsilon$ is  0.001, the fluctuations do not affect the 
entanglement.  
But at higher mean values, fluctuations in $\epsilon$ affect entanglement 
dynamics.  Taking mean of $\epsilon$ to be $0.3$, the time-evolution of $Y$ 
is shown 
in Fig. \ref{fig:fluc} for $\lambda=0.001$ and $0.1$.  The fluctuating driving 
term is taken to be piecewise constant function.  The total evolution time is 
divided into 100 equal parts;  in each  part, the value of $\epsilon$ is 
chosen at random from the Gaussian distribution defined earlier.  
The evolution is tracked for ten different random trials in each case.  
The evolution of $Y$ corresponding to small $\lambda$, 
shown in (a) of Fig. \ref{fig:fluc}, shows a visible spread in the evolution 
during different trials.  In order to distinctly bring out the effects of 
fluctuations, the dynamics is continued for five units of scaled time.  
The entanglement itself is very small as 
the field-field coupling is small. But the fluctuations induce large 
relative changes in the entanglement.  Thus, strong pumping and weak 
coupling between the cavities lead to large uncertainties in entanglement.      
 With larger $\lambda$, refer (b) in Fig. 
\ref{fig:fluc},   the spread is almost insignificant.   Strong coupling  
between the modes reduces the effect of the  driving-field fluctuations on the 
entanglement.  As stressed earlier, sufficient coupling between the cavities 
allows for efficient  flow of energy and the maximum value 
of $Y$ during evolution is also large.  It is, therefore, possible to choose  
the cavity-cavity coupling so that the effects of fluctuations are minimized. 
In other words, smaller coupling between the cavities necessitates a more stable 
pumping. 

\section{Summary}
Two coupled cavities generate entangled, bipartite, non-Gaussian states.  
In the absence of external pumping, the cavities exchange energy periodically. 
The entanglement attained during evolution increases with the number of photons 
present in the cavities  initially. Inclusion of external pumping in one of the 
cavities amplifies the field in that cavity, which, in turn, energizes, 
{\em via} the evanescent coupling, the mode in the other cavity.   This affects 
the entanglement between the modes.  The covariance criterion of entanglement 
is insensitive to the driving amplitude if the driving is linear.   However, 
with quadratic pumping, the states 
occurring during the evolution are such that the  covariance criterion 
identifies  entanglement.  If the couping between cavities is weak, increasing 
the driving field amplitude decreases the peak entanglement attainable during   
evolution compared to the pump-free evolution.  Therefore, it is important to 
tailor the pumping strength to match the cavity-cavity coupling.   During 
evolution, maximum entanglement between the fields in the cavities occurs 
when the mean number of photons in the two cavities are equal.\\

  For weak coupling between the cavities,  the maximum attainable entanglement 
decreases rapidly as the pumping strength is increased. Though the entanglement 
increases for weak pumping, it begins to decrease for further increase of pumping 
strength.   This is because of preferential increase in the energy of the pumped 
cavity leading to large difference in the mean number of photons of the two 
cavities.  On the other hand, if the coupling between the cavities is stronger, 
the maximum entanglement attained during evolution increases for larger range of 
values of the pumping strength.  Also, the change of entanglement with pumping  
strength is not as rapid as the change if the coupling between the cavities is 
weak.   Consequently, the evolution of entanglement is less sensitive to 
fluctuations in the pumping  if the cavities are coupled effectively.   

\newpage
\bcen
{\bf Figure Captions}
\ecen
\noindent
Fig.1 Variation of $Y$ with time.  The time-axis represents scaled-time 
defined as the ratio of time to $\pi/\lambda$.  One unit of scaled-time 
corresponds to $\pi/\lambda$. The curves shown correspond to different 
total quantum numbers: N=1 
(dash), 5 (dot-dash), 10 (continuous) and 50 (dot).  In all cases the coupling 
$\lambda$ is 0.1.\\\\
Fig. 2 Variation of covariance measure $Y$ (continuous)  and 
von Neumann entropy $S$ (dash) with scaled time. The initial state 
is $\vert 5,0\ra$ and $\lambda =0.1$.\\\\
Fig. 3 Comparison of evolution of $Y$ for weak and strong couplings when 
pumping is quadratic.  
Different plot types are used for different combinations of $\lambda$ and 
$\epsilon$: dashed - $\lambda=0.001$ and $\epsilon=0.1$;   
continuous - $0.001,0.001$, beaded - $0.1,0.1$. The continuous curve 
hides the curve corresponding to the case: $\lambda=0.1$ and $\epsilon =
0.001$.  The initial state is $\vert 5,0\ra$.\\\\
Fig. 4 Ratio of $\vert\la\ano-\bno\ra\vert$ to $\la\ano+\bno\ra$ as a 
function of time.  The curves shown 
correspond to $\lambda=0.001$ and $\epsilon=0.1$ (dash);   
$0.001,0.001$ (continuous), $0.1,0.005$ (dots). 
\\\\
Fig. 5 Maximum value of $Y$ as a function of $\epsilon$.  
For weak coupling $\lambda=0.001$  
(continuous)and for strong coupling  $\lambda=0.05$ (dots).  Additional 
curves correspond to $\lambda = 0.005$ (dot-dash) and 0.01 (dash).  
The initial state is $\vert 5,0\ra$.\\\\
Fig. 6 Evolution of $Y$ in time under fluctuating driving field.  
The mean driving amplitude is 0.3.  
Two values of field-field couplings are assumed: 
$\lambda=0.001$ [figure (a)]  and $\lambda=0.05$ [figure(b)].  Different 
evolution profiles correspond to different trials.  In each of the figures, 
results of ten trial evolutions are given. 
The initial state is $\vert 5,0\ra$ for all the profiles.

\newpage
\end{document}